\documentstyle[12pt,epsfig]{article}
\begin{document}
\title{Effects of Massive Gluons on Quarkonia \\ in Light-Front QCD}
\author{Martina M. Brisudov\'a $^2$, S\'ergio Szpigel $^1$, and Robert J. Perry
$^1$\\ \\{$^1$ \it Department of Physics}\\ { \it The Ohio State University,
Columbus, OH 43210}\\  \\{$^2$ \it Theoretical Division}\\ {\it Los Alamos National Laboratory, Los Alamos, NM 87545}}    

\date{\today}
\maketitle
\abstract{A constituent parton picture of hadrons with logarithmic 
confinement arises naturally in light-front QCD when the hamiltonian is
computed using a perturbative renormalization group; however, rotational
symmetry is not manifest as a simple kinematic symmetry.  Relevant operators
must typically be fine-tuned when a perturbative renormalization group is
used, and in light-front field theory such operators contain functions of
longitudinal momenta.  We explore the possibility that a gluon mass operator
with simple longitudinal momentum dependence may improve approximate
kinematic rotational symmetry, as revealed by the potential between heavy
quarks.}
\vfill
\eject

\vskip .25in

The solution of Quantum Chromodynamics in the nonperturbative domain remains
one  of the most important problems in physics. A nonperturbative approach based 
on light-front quantization has been suggested by Wilson et al \cite{thelongpaper}. 
The basic assumption is  that it is possible to {\it derive a constituent picture 
for hadrons from QCD} \cite{thelongpaper,P2,c7}. If this is possible, nonperturbative 
bound state  problems in QCD can be approximated as coupled, few-body Schr{\"o}dinger 
equations.  

To arrive at a constituent approximation, vacuum fluctuations need to be 
isolated. This is achieved by formulating the theory in light-front
coordinates, in which the vacuum is made trivial by the introduction of a
longitudinal momentum cutoff and vacuum effects must be replaced by effective
interactions. However, a trivial vacuum alone does not ensure the emergence of a 
constituent picture. There may still  be  infinitely many Fock components
needed to describe a hadron. In addition to removing vacuum components from the
states, significant coupling between few-body and many-body states must be
replaced by effective interactions involving few-body states alone. This can be
achieved, if the many-body states have high energy, by a renormalization group
\cite{lfrenorm}.

We sketch the simplest version of this procedure and motivate the current
calculations.  The first step is to use a similarity renormalization group
\cite{similarity} and coupling coherence \cite{coupcoh} to compute the
renormalized hamiltonian to a fixed order in the running canonical coupling. 
The truncation of this expansion introduces errors in the strengths of all
relevant, marginal, and irrelevant operators; all of which contain functions of
longitudinal momenta because longitudinal locality is not maintained in
light-front field theory \cite{coupcoh}.  It has been shown that the
hamiltonian computed to second order contains a two-body logarithmic confining
interaction that acts in every Fock sector to confine quarks and gluons
\cite{P2}.  This interaction alone is not rotationally invariant, and may
require rotational symmetry to emerge dynamically only after different Fock
sectors mix in the eigenstates. 

The largest errors in the final results may well originate from this
perturbative approximation of the effective hamiltonian. That such an
approximation may not be absurd is indicated by lattice gauge theory, where
perturbative renormalization and tadpole resummation leads to extremely
accurate results for heavy quark systems \cite{lattice}. Since errors in the
relevant operators exponentiate in the scaling regime, it is likely that these
operators must be fine-tuned.  In lattice field theory one avoids most relevant
operators by maintaining gauge invariance, although a relevant operator must be
tuned to restore chiral symmetry if Wilson fermions are used.  The relevant
operators of immediate interest to us are quark and gluon self-energies, and in
this paper we focus on the possibility that fine-tuning the gluon self-energy
may be required to restore rotational invariance.

The second step in our approach is solving for the hadronic 
bound states. If one tries to diagonalize the entire effective Hamiltonian, 
the wave function of a hadron must contain arbitrarily many parton components 
because the effective Hamiltonian still contains emission and absorption,
albeit  energetically limited. Instead, we divide the effective Hamiltonian
into a part, $H_0$, which is solved nonperturbatively, and the remaining part,
$V$, that is treated in bound state perturbation theory (BSPT). The division is
arbitrary,  but the choice of $H_0$ determines the convergence of the
calculation, and we must choose $H_0$ to approximate the physics relevant
for hadronic bound states.

If a constituent approximation emerges, we may hope to employ only interactions
that preserve parton number in $H_0$.  This is not required in our approach,
since it is possible that mixing between few-parton sectors is non-perturbative
without requiring many-parton sectors.  If it is possible to completely ignore
emission and absorption of partons in $H_0$, mesons can be approximated as
quark-antiquark bound states with no gluons.  However, rotational symmetry is
maintained in this case only if the quark-antiquark potential is rotationally
symmetric.  Of course, we expect excited states in which the pair is separated
by a large distance to contain gluons, so that {\it it is only the short and
intermediate range parts of the potential that need to be rotationally
symmetric}. Even in these ranges the potential need only be approximately
invariant, since many-body effects found in higher orders of BSPT are required
to restore complete rotational symmetry, as is well illustrated by light-front
QED calculations \cite{trulko}.

In previous work \cite{c7,us} a nonrelativistic reduction of the second-order 
effective Hamiltonian was used to study mesons containing at least one heavy 
quark, approximated in leading order as color-singlet $q\bar{q}$ bound states.
$H_0$ included the short range Coulomb interaction and a rotationally invariant
long-range logarithmic potential obtained by angular averaging the long-range
potential in the hamiltonian.  $V$ included a non-rotationally invariant
long-range potential, as well as low energy gluon emission and absorption, and
other operators.  The non-rotationally invariant potential leads to corrections
in first-order BSPT, reported in \cite{c7} to be of order 30\%. However, a
mistake was found in the calculation of the non-rotationally invariant term.
The calculation with the corrected potential reveals that the corrections are
actually  less than 10\%. Such corrections are not necessarily a serious
problem, because second-order BSPT involving gluon emission and absorption will
also lead to corrections of the same order when we use couplings
required to produce heavy meson spectra.  However, the calculation of these
second-order BSPT corrections requires the solution of a three-body problem
with confining two-body interactions, and we have not yet attempted this
calculation.  Since there are several ways in which rotational symmetry may
emerge, we have chosen to investigate the simplest possiblity in this paper.

The QCD hamiltonian contains quark and gluon self-energies that depend on
longitudinal momenta even at second order.  These self-energies and the
logarithmic confining interaction produce a mass gap between the $q\bar{q}$ and
$q\bar{q}g$ sectors, making a constituent approximation plausible.  The
self-energies are relevant operators, and the second-order approximation is
probably inadequate; so we assign low- and  intermediate-energy gluons an
effective mass $M_g(x_g)$, and partially  integrate them out. Note that we
must allow this operator to be a function of the gluon longitudinal  momentum
fraction, $x_g$. This leads to new effective  interactions in the Hamiltonian.
The functional form of $M_g(x_g)$ is determined  by the requirement that the
violation of rotational symmetry in the effective  interactions is reduced. We
have found that this is achieved by using a $x_g$-dependent effective gluon
mass of the form $M_g^2(x_g)=m_g ^2 x_g^2$. No  other power of $x_g$, including
a constant, or a polynomial in which the lowest  term is not $x_g^2$, leads to
such an effect. By kinematical considerations, the parameter $m_g$ is expected
to be of order the glueball mass.\footnote{Note that $m_g$ is not one-half the
glueball mass because of the $x$-dependence in this operator.} In this letter
we present  numerical results obtained by applying the approach to heavy
quarkonia for which the approximations are best justified.   
 
In previous applications \cite{P2,c7,us,trulko} the similarity transformation
has been performed perturbatively around the free Hamiltonian. The starting
point is a  ``bare" QCD hamiltonian, $H^{B}_{\Lambda}$, in the light-front gauge 
($A^+=0$) regulated by a large cutoff , $\Lambda$, on energy differences at
the interaction  vertices. Counterterms must be added to this, $H^{c.t.}_{\Lambda}$; 
and coupling-coherence fixes all couterterms. The transformed Hamiltonian $H_{\lambda}$ 
is expressed as
\begin{eqnarray}
H_{\lambda} = S H_{\Lambda} S^{\dagger}=S \left(H^{B}_{\Lambda}+ H^{c.t.}_{\Lambda}\right)S^{\dagger},
\end{eqnarray}
where  $S$ is a unitary transformation ($S^{\dagger}S=1$) such that
the  Hamiltonian $H_{\lambda}$ is band diagonal with respect to the scale
$\lambda$,  i.e.,
\begin{equation}
H_{\lambda} = f_{\lambda} G_{\lambda} \;, 
\end{equation}
where $f_{\lambda}$ is a cutoff 
function which makes the Hamiltonian vanish outside the band. Equivalently,
\begin{eqnarray}
{d H_{\lambda} \over{d\lambda}} = \left[ H_{\lambda}, T_{\lambda}\right]\;,
\end{eqnarray}
where
\begin{eqnarray}
T_{\lambda} = S_{\lambda}^{\dagger} {dS_{\lambda}\over{\lambda}} \ \  .
\end{eqnarray}
Let $\lambda^2$ be the transverse cutoff (with dimension of $mass^2$), ${\cal
P}^+$ a longitudinal scale (needed for  dimensional reasons), and let  $E_0$
denote eigenvalues of the free light-front Hamiltonian $h$. We can choose the
cutoff function $f_{\lambda_{ij}}= \theta\left( {\lambda^2\over{{\cal P}^+}}
-\vert E_{0i}-E_{0j}\vert\right)$, for example, which fixes the transformation
in Eq.$(3)$. We then use Eq. (3) to determine how all operators in the
Hamiltonian transform, and coupling coherence fixes the initial strengths of
all non-canonical operators in terms of the canonical coupling, leading finally
to an expansion for the renormalized Hamiltonian $H^{R}_{\lambda}$ in powers of
the canonical coupling.

A perturbative error is made in the strength of each operator, and this error
typically grows as $\lambda$ decreases because the canonical coupling grows. 
As stated above, errors in the strength of relevant operators are expected to
produce the largest bound state errors, and we assume that at some scale these
errors become sufficiently large that we must fix them by adding mass terms
that are tuned by hand.  We will add a gluon mass term at a scale $\lambda$,
which from this point will refer only to the specific cutoff where the gluon
mass is added.  Moreover, since the gluon mass makes a significant contribution
to the diagonal components of the Hamiltonian, we alter the transformation so
that we now perturb about the free Hamiltonian with the gluon mass included. 
The first step is to lower the cutoff to the scale $\lambda$ and obtain the
renormalized Hamiltonian,
\begin{equation}
H^{R}_{\lambda}=h+f_{\lambda}{\bar v}_{\lambda}. 
\end{equation}

At this point we add an effective $x$-dependent gluon mass operator to
$H^{R}_{\lambda}$ and regroup the Hamiltonian so that the new one-body operator
is included in the free Hamiltonian
\begin{equation}
H^{R}_{\lambda}\rightarrow H^{\prime R}_{\lambda}=h^{\prime}+f_{\lambda}{\bar 
v}_{\lambda}\;,
\end{equation}
\begin{equation}
h^{\prime}=h+\sum_{i} \int_{q}\frac{M_g^2(x_g)}{q^+} a_{i,q}^\dagger a_{i,q}  .
\end{equation}
Below the scale $\lambda$ the new dispersion relation for gluons is given
by:  
\begin{eqnarray}
q^{-}=\frac{{q_\perp}^2}{q^+}\rightarrow \frac{{q_\perp}^2+M_g^2(x_g)}{q^+}\;.
\end{eqnarray} 
\noindent
Now, we run a second similarity transformation. Since we have changed the free 
Hamiltonian, we have to use a similarity transformation that runs a cutoff for 
the new gluon dispersion relation. This second transformation is implemented by 
introducing a new similarity function, ${\tilde f}_{\lambda}$.

At this point the Hamiltonian is band diagonal with respect to the cutoff
function $f$, but not with respect to the cutoff function ${\tilde f}$.  We
cannot simply change the cutoffs by hand, so we must run the second
transformation from infinity to whatever final value of the cutoff we choose.
The procedure for running this second transformation is formally the same as
that for running the first transformation. There
is no need for new counterterms, since we have already completely
renormalized the hamiltonian and the second transformation does not  generate
any new divergences. The output of the second transformation is an 
${\cal O}(g^2)$ effective Hamiltonian band-diagonal in light-front energy with 
respect to the scale ${\mu^2\over{{\cal P}^+}}$, and can be written as :
\begin{eqnarray}
{\tilde H}^{R(2)}_{\mu}&=&h^{\prime}+{\tilde f}_{\mu}{\bar 
v}^{(2)}_{\mu}\nonumber\\
&=&h^{\prime} +  v^{(1)}  +  v^{(2)}  +   
v^{(2)}_{eff} + {\tilde v}^{(2)}_{eff} \;.
\end{eqnarray}

The light-front kinetic energy is augmented by the effective gluon mass
operator, and it is still diagonal. The ${\cal O}(g)$ term $ v^{(1)}$ now 
corresponds to (massive) gluon emission and absorption  with nonzero matrix 
elements only between states with energy difference smaller than 
${\mu^2\over{{\cal P}^+}}$. The gluon mass sets a dynamical threshold for gluon 
emission and  absorption. Since the mass of a free gluon can be arbitrarily
small depending on the $x_g$-dependence chosen, particle number changing
interactions are, in  general, not integrated out completely. This means that
there still may be a contribution from the $q\bar{q}g$ bound state, but
compared to the previous work where the  low-energy gluons were left massless
in the effective Hamiltonian, this contribution is significantly reduced. The 
${\cal O}(g^2)$ canonical  instantaneous gluon exchange interaction, $v^{(2)}$,
and the  ${\cal O}(g^2)$  effective interactions generated by the first
transformation, $v^{(2)}_{eff}$,  remain unchanged, apart from the new overall
similarity function that makes the Hamiltonian band diagonal with respect to
the scale $\mu$. The new ${\cal  O}(g^2)$ effective interactions, ${\tilde
v}^{(2)}_{eff}$, generated by the second transformation depend on the gluon
mass $m_g$ and its functional form.

Using $M_g^2(x_g)=m_g^2 x_g^2$ the effective $q\bar{q}$  potential in 
the nonrelativistic limit, including the new terms is given by:
\begin{eqnarray}
v_{q\bar{q}}&\rightarrow &
\theta\left(\frac{\mu^2}{{\cal P}^+}-\left|E_{0i}-E_{0j}\right|\right) 4g^2 
C_{F}(2M)^2 \left[-\frac{1}{{\bf q}^2} -\frac{1}{q_z^2}\frac{q_{\perp}^2}{\bf 
q^2}
 \theta\left(\frac{\lambda^2}{{\cal P}^+}-\frac{{\bf q}^2(2M)}{|q_z| 
P^+}\right)\right.\nonumber\\
&+& \left.\frac{1}{q_z^2}\frac{q_{\perp}^2}{{\bf q}^2+\frac{m_g^2}{(2M)^2} 
q_z^2}
\theta\left(\frac{\lambda^2}{{\cal P}^+}-\frac{{\bf q}^2 (2M)}{|q_z| 
P^+}\right)\theta\left(\frac{({\bf q}^2+\frac{m_g^2}{(2M)^2} q_z^2) 2M}{|q_z| 
P^+}-\frac{\mu^2}{{\cal P}^+}\right)\right] \;,
\label{pot}
\end{eqnarray}
where ${\bf q} \equiv (q_z,q_{\perp})$ is the exchanged (gluon) momentum, $M$ 
is the quark mass, $C_F=4/3$ is a color factor and $P^+$ is the meson total 
longitudinal momentum. To take the nonrelativistic limit, we introduce a 
$z$-component for momenta,
\begin{equation}
p^+ = \sqrt{{\bf p}_\perp^2+p_z^2+M^2}+p_z \;,
\end{equation}
so that $q_z = p_{1z}-p_{2z}$, and we make an expansion in powers of $p/M$,
dropping all non-leading terms. 

The potential in eqn. $(10)$ results from instantaneous gluon exchange in the 
light-cone gauge added to terms from the two similarity transformations. 
Each similarity transformation produces a second-order gluon exchange interaction 
that replaces direct exchange of gluons not allowed by the cutoffs on energy change 
at the vertices. If we had not added a gluon mass and run a second transformation, 
the parameters would be the cutoff $\mu$, the quark mass $M$, and the
coupling $\alpha$. The second similarity transformation introduces two 
additional parameters, $\lambda$ and $m_g$.   

The purpose of this work is to explore the possibility that the gluon mass
operator may lead to an effective quark-antiquark potential that is more
nearly rotationally invariant than the potential computed in our previous
work. To show this,  we choose $H_0$ as in our previous work (i.e. the 
nonrelativistic reduction of  the kinetic energy, the effective one-body 
operators, Coulomb potential and the angular average of the confining
potential with constituent masses). The bound state calculation proceeds in
full analogy with \cite{c7,us}. We assume that the scale $\mu $ is small enough
that the lowest-lying states are dominated by their $q\bar{q}$ component.

The potential given in Eq. ~\ref{pot} is Fourier transformed and expanded  in
even Legendre polynomials,
\begin{equation}
V({\bf r})=V_{\rm coul}(r)+\sum_{k=0}^{\infty}V_{\rm conf, 2k}(r)P_{\rm 2k}(z)
\; ,
\end{equation}
where $r$ is the quark-antiquark separation, $z=cos(\theta)$ and
\begin{equation}
V_{\rm conf,2k}(r)=\frac{4k+1}{2} \int_{-1}^{+1} dz \hspace{0.2cm} V({\bf
r})P_{\rm 2k}(z) \; .
\end{equation}
The eigenvalue problem for the leading order hamiltonian $H_0$ is
\begin{eqnarray}
H_{0} \vert P \rangle = {\cal M}^2 \vert P \rangle \  \  , 
\end{eqnarray}
where ${\cal M}^2$ is the invariant mass of the bound state and ${\cal M}^2 = 4
M^2+4 M E$ defines $E$. This leads to a  dimensionless  Schr\"odinger equation
\cite{us}: 
\begin{eqnarray}
\left[ -{d^2\over{d{\vec{\cal R}}^2}}
+ c \left(
 {1\over{\pi}}V_{\rm conf,0}({\cal R}) + V_{\rm coul}(
{\cal R})\right)\right] \psi({{\vec{\cal R}}}) =
 e \psi({{\vec{\cal R}}}) \  \  ,
\end{eqnarray}
with
\begin{eqnarray}
{\cal R} \equiv  {\cal L}r , \hspace{0.5cm} c \equiv {2m\alpha C_F \over{{\cal 
L}}} , \hspace{0.5cm} e & \equiv & {2m (E-\tilde{\Sigma}) \over{{\cal L}^2}} ,
\end{eqnarray}
\noindent 
where ${\cal L} \equiv (\mu^2/{\cal P}^+)(P^+/2M)$ , $m=M/2$ is the reduced 
mass, $\alpha=g^2/4 \pi$ and $\tilde{\Sigma }$ is the finite shift produced by 
the self-energies after subtracting terms needed to make the confining potential vanish at the origin,
\begin{eqnarray}
{\tilde \Sigma}&=&
\frac{\alpha C_F {\cal L}}{\pi}\frac{\lambda^2}{\mu^2}
\left[\left(1+\frac{3M}{2{\cal L}\frac{\lambda^2}{\mu^2}}\right)
{\rm ln}\left(\frac{M}{{\cal L}\frac{\lambda^2}{\mu^2}+M}\right)
+\frac{1}{4}\frac{M}{{\cal L}\frac{\lambda^2}{\mu^2}+M}+\frac{1}{4}+\frac{2\mu^2}{\lambda^2}\right. \nonumber\\
&&~~~~~~~~+\left.\frac{\mu^2}{\lambda^2}{\rm ln}\left(\frac{2M}{\cal L}\right) -{\rm ln} 
\left(\frac{2M \mu^2}{{\cal L}\lambda^2}\right)+\left(1+\frac{4 M^2}{m_g^2} - \frac{\mu^2}{\lambda^2}\right)
{\rm ln}\frac{4 M^2}{m_g^2+4 M^2}
\right]\nonumber\\
&-& \frac{\alpha C_F M}{2\pi} \int_{0}^{1}dy \int_{0}^{\infty} dw 
\hspace{0.2cm}\theta\left(\frac{{\cal L}\lambda^2}{M \mu^2}
-\frac{(y^2 +w)}{y (1-y) }\right)
\theta\left(\frac{ (y^2 + w)}{y (1-y) }
+\frac{m_g^2 y^2}{4 M^2 y }-\frac{{\cal L}}{M}\right)\nonumber\\
&&~~~~~~~~\times \left[\frac{2w}{y^2}
+\frac{w}{1-y}+\frac{y^2 }{1-y}\right]
\frac{1}{y^2 +w+y^2 (1-y) m_g^2/(4M^2)}\;.
\end{eqnarray}

The parameters are adjusted so that the angular average of the potential is 
close to the one used for charmonium in our previous work, as displayed in Fig. 1. 
For a qualitative analysis this is sufficient. Here, the potential obtained in ref. 
\cite{c7} is calculated by using $M=1.6$ GeV, $\mu=1.8$ GeV and 
$\alpha=0.53$.\footnote{An error in the expression for the self 
energy in ref. \cite{c7} has been corrected. This changes the set of parameters
required to fit the lowest lying states slightly from those in \cite{c7}.}     
For the new potential we use $M=1.6$ GeV, $m_g=1.6$ GeV, $\lambda=9.5$ GeV, 
$\mu=1.5$ GeV and 
$\alpha=0.55$ to obtain a reasonable fit for the masses of the lowest lying
states of charmonium: ${\cal M}_{1S}=3.1$ GeV, ${\cal M}_{1P}=3.5$ GeV and ${\cal
M}_{2S}=3.6$ GeV. 

In Fig. 2 we  compare the angular dependence of the full potential to the one
obtained in our previous calculation.  The violation of  rotational symmetry in
the new effective potential is significantly reduced in the intermediate range
($0.3-1.5$ fm) and increases at large distances. As stated above, the long
range part of the potential need not be rotationally invariant because states
in which the $q\bar{q}$ pair are separated by large distances will contain
valence gluons.  At short distances ($<0.3$ fm), where the Coulomb potential
dominates, the violation is very small, although larger for the new potential.
With the parameters fixed we evaluate the corrections due to  rotationally
noninvariant terms in the confining potential. There are  corrections to
P-states in first-order BSPT, while S-states are corrected in second-order. 
For weak couplings these corrections scale with the coupling at the same order
as second-order BSPT corrections involving gluon emission and absorption. We
concentrate on the corrections to the lowest lying P-state, which are easily
computed.  These corrections to the lowest lying P state of heavy quarkonium are 
reduced from about 8\% \cite{c7} to about 2\%.
\vspace{0.5cm}    

In conclusion, we have shown that it is possible to adjust the relevant gluon
mass operator so that apparent violations of rotational invariance in heavy
quarkonia are reduced.  It is assumed that non-perturbative corrections to this
operator become important as the hadronic bound state scale is approached by
the cutoff, and that fine-tuning this operator allows us to perturbatively
reduce the cutoff further without modifying other operators by hand.  We use
rotational invariance as a test of these ideas, tuning the dependence of the
gluon mass on longitudinal momentum to reduce naive violations of rotational
invariance by a factor of $3-4$.  Rotational invariance of the two-body
interaction is significantly improved at intermediate distances if a gluon mass
operator of the form $m_g^2 x_g^2$ is used. Corrections to rotational invariance 
are sufficiently small in charmonium even without such a gluon mass that we can not 
yet conclude that this operator is required. A real test of the significance of the 
relevant gluon mass operator requires higher order BSPT calculations and a study 
of light mesons.

We would like to thank Ken Wilson, Terry Goldman, Brent Allen, Billy Jones and 
Stan Brodsky for discussions. S.S. is a CNPq-Brazil fellow (proc. 204790/88-3). 
The work of R.P. and S.S. was supported by National Science Foundation grant 9511923. 
The work of M.B. was supported by the United States Department of Energy.

\newpage
\section*{Figure Captions}
{\bf Figure 1:}
The angular average of confining and Coulomb potential with massive gluons 
compared to the angular average with massless gluons. The parameters of the 
potential with massive gluons were adjusted so that it is close to the 
potential with massless gluons.

\vspace{0.5cm}
\noindent
{\bf Figure 2:}
The angular dependence of the potential with massive gluons compared to
 the potential with massless gluons at $r=0.2$ fm, $1.0$ fm, $2.0$ fm. The violation 
of rotational symmetry is significantly reduced in an intermediate range ($0.3 -1.5$ fm). 
At very short distances the violation of rotational symmetry is small, although larger for 
the potential with massive gluons.

\newpage
\begin{figure}[ht]
\begin{center}
\epsfig{figure=figure1.eps}
\vspace{2cm}
\caption{The angular average of confining and Coulomb potential with massive gluons compared to 
the angular average with massless gluons. The parameters of the potential with massive gluons 
were adjusted so that it is close to the potential with massless gluons.}
\end{center}
\end{figure}
\newpage
\begin{figure}[ht]
\begin{center}
\epsfig{figure=figure2.eps}
\vspace{2cm}
\caption{The angular dependence of the potential with massive gluons compared to the potential 
with massless gluons at $r=0.2$ fm, $1.0$ fm, $2.0$ fm. The violation of rotational symmetry is 
significantly reduced in an intermediate range ($0.3 -1.5$ fm). At very short distances the violation 
of rotational symmetry is small, although larger for the potential with massive gluons.}
\end{center}
\end{figure}

\end{document}